\documentclass[letterpaper,peer-reviewed]{aes135}
\usepackage[utf8]{inputenc}
\usepackage{amsmath}
\usepackage{amssymb}
\usepackage{graphicx}

\makeatletter

\usepackage{url}
\authors{Jean-Marc Valin\aff{1}, 
        Gregory Maxwell\aff{1}, 
        Timothy B. Terriberry\aff{1}, 
         and Koen Vos\aff{2}}

\affiliation[1]{Mozilla, Xiph.Org}
\affiliation[2]{Microsoft}

\correspondence{Jean-Marc Valin}{jmvalin@jmvalin.ca}

\lastnames{Valin, Terriberry, Vos}

\title{High-Quality, Low-Delay Music Coding in the Opus Codec}

\shorttitle{Music Coding in Opus}
\newcommand{\itemspace}{\vspace{-1.2mm}}

\begin{abstract}
The IETF recently standardized the Opus codec as RFC6716. Opus targets a wide range of real-time Internet applications by combining a linear prediction coder with a transform coder. We describe the transform coder, with particular attention to the psychoacoustic knowledge built into the format. The result out-performs existing audio codecs that do not operate under real-time constraints.
\end{abstract}

\makeatother

\begin{document}
\maketitle

\section{Introduction}

In RFC~6716~\cite{rfc6716} the IETF recently standardized Opus~\cite{Opus-website},
a highly versatile audio codec designed for interactive Internet applications.
This means support for speech and music, operating over a wide range
of changing bitrates, integration with the Real-Time Protocol (RTP),
and good packet loss concealment, with a low algorithmic delay. 

Opus scales to delays as low as 5~ms, even lower than AAC-ELD (15~ms).
Applications such as network music performance require these ultra-low
delays~\cite{Carot-thesis}. Despite the low delay, Opus is competitive
with high-delay codecs designed for storage and streaming, such as
Vorbis and the HE and LC variants of AAC, as the evaluations in Section~\ref{sec:Evaluation-and-Results}
show.

Opus supports
\begin{itemize}
\item \itemspace Bitrates from 6 kb/s to 510 kb/s,
\item \itemspace Five audio bandwidths, from narrowband (8~kHz) to fullband
(48~kHz),
\item \itemspace Frame sizes from 2.5 ms to 60 ms,
\item \itemspace Speech and music, and
\item \itemspace Mono and stereo coupling.
\end{itemize}
\itemspace In-band signaling can dynamically change all of the above,
with no switching artifacts.

We created the Opus codec from two core technologies: Skype's SILK~\cite{SILK-draft-02}
codec, based on linear prediction, and Xiph.Org's CELT codec, based
on the Modified Discrete Cosine Transform (MDCT). Section~\ref{sec:Overview-of-Opus}
presents a high-level view of their unification into Opus. After many
major incompatible changes to the original codecs, the result is open
source, with patents licensed under royalty-free terms\footnote{http://opus-codec.org/license/}.
The reference encoder is professional-grade, and supports
\begin{itemize}
\item \itemspace Constant bit-rate (CBR) and variable bit-rate (VBR) rate
control,
\item \itemspace Floating point and fixed-point arithmetic, and
\item \itemspace Variable encoder complexity.
\end{itemize}
\itemspace Its CBR produces packets with exactly the size the encoder
requested, without a bit reservoir to imposes additional buffering
delays, as found in codecs such as MP3 or AAC-LD. It has two VBR modes:
Constrained VBR (CVBR), which allows bitrate fluctuations up to the
average size of one packet, making it equivalent to a bit reservoir,
and true VBR, which does not have this constraint.

This paper focuses on the CELT mode of Opus, which is used primarily
for encoding music. Although it retains the fundamental principles
of the original algorithms published in~\cite{Valin2009,Valin2010}
reviewed in Section~\ref{sec:Constrained-Energy-Lapped-Transform},
the CELT algorithm used in Opus differs significantly from that work.
We have psychoacoustically tuned the bit allocation and quantization,
as Section~\ref{sec:Quantization-and-Encoding} outlines, and designed
additional tools to conceal artifacts, which Section~\ref{sec:Psychoacoustic-Improvements}
describes.

\section{Overview of Opus}

\label{sec:Overview-of-Opus}
\begin{figure}
\centering{\includegraphics[width=1\columnwidth]{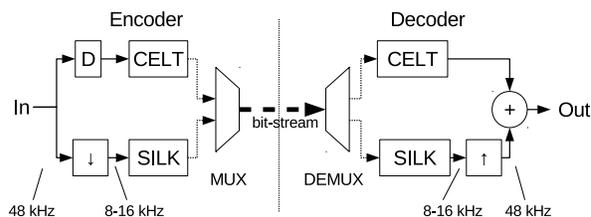}}

\caption{High-level overview of the Opus codec.\label{fig:High-level-overview-of-Opus}}
\end{figure}

Opus operates in one of three modes:
\begin{itemize}
\item \itemspace SILK mode (speech signals up to wideband),
\item \itemspace CELT mode (music and high-bitrate speech), or
\item \itemspace Hybrid mode (SILK and CELT simultaneously for super-wideband
and fullband speech).
\end{itemize}
\itemspace Fig.~\ref{fig:High-level-overview-of-Opus} shows a high-level
overview of Opus. CELT always operates at a sampling rate of 48~kHz,
while SILK can operate at 8~kHz, 12~kHz, or 16~kHz. In hybrid mode,
the crossover frequency is 8~kHz, with SILK operating at 16~kHz
and CELT discarding all frequencies below the 8~kHz Nyquist rate.

CELT's look-ahead is 2.5~ms, while SILK's look-ahead is 5~ms, plus
1.5~ms for the resampling (including both encoder and decoder resampling).
For this reason, the CELT path in the encoder adds a 4~ms delay.
However, an application can restrict the encoder to CELT and omit
that delay. This reduces the total look-ahead to 2.5~ms. 

\subsection{Configuration and Switching}

Opus signals the mode, frame size, audio bandwidth, and channel count
(mono or stereo) in-band. It encodes this in a \emph{table-of-contents}
(TOC) byte at the start of each packet~\cite{rfc6716}. Additional
internal framing allows it to pack multiple frames into a single packet,
up to a maximum duration of 120~ms. Unlike the rest of the Opus bitstream,
the TOC byte and internal framing are not entropy coded, so applications
can easily access the configuration, split packets into individual
frames, and recombine them.

Configuration changes that use CELT on both sides (or between wideband
SILK and Hybrid mode) use the overlap of the transform window to avoid
discontinuities. However, switching between CELT and SILK or Hybrid
mode is more complicated, because SILK operates in the time domain
without a window. For such cases, the bitstream can include an additional
5~ms \emph{redundant CELT frame} that the decoder can overlap-add
to bridge and gap between the discontinuous data. Two redundant CELT
frames---one on each side of the transition---allow smooth transitions
between modes that use SILK at different sampling rates. The encoder
handles all of this transparently. The application may not even notice
that the mode has changed.

\section{Constrained-Energy Lapped Transform (CELT)}

\label{sec:Constrained-Energy-Lapped-Transform}
\begin{figure}
\includegraphics[width=1\columnwidth]{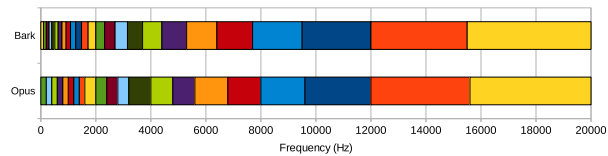}

\caption{CELT band layout vs. the Bark scale.\label{fig:CELT-band-layout}}
\end{figure}

Like most transform
coding algorithms, CELT is based on the MDCT. However, the fundamental
idea behind CELT is that the most important perceptual aspect of an
audio signal is the spectral envelope. CELT preserves this envelope
by explicitly coding the energy of a set of bands that approximates
the auditory system's critical bands~\cite{Moore}. Fig.~\ref{fig:CELT-band-layout}
illustrates the band layout used by Opus. The format itself incorporates
a significant amount of psychoacoustic knowledge. This not only reduces
certain types of artifacts, it also avoids coding some parameters. 

\begin{figure}
\centering{\includegraphics[width=1\columnwidth]{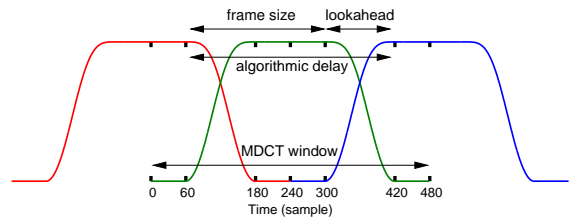}}

\caption{Low-overlap window used for 5~ms frames.\label{fig:Low-overlap-window}}
\end{figure}

CELT uses flat-top MDCT windows with a fixed overlap of 2.5~ms, regardless
of the frame size, as Fig.~\ref{fig:Low-overlap-window} shows. The
overlapping part of the window is the Vorbis~\cite{VorbisSpec} power-complementary
window
\begin{equation}
w(n)=\sin\left[\frac{\pi}{2}\sin^{2}\left(\frac{\pi\left(n+\frac{1}{2}\right)}{2L}\right)\right]\ .
\end{equation}
Unlike the AAC-ELD low-overlap window, the Opus window is still symmetric.
Compared to the full-overlap of MP3 or Vorbis, the low overlap allows
lower algorithmic delay and simplifies the handling of transients,
as Section~\ref{sub:Handling-of-Transients} describes. The main
drawback is increased spectral leakage, which is problematic for highly
tonal signals. We mitigate this in two ways. First, the encoder applies
a first-order pre-emphasis filter $A_{p}\left(z\right)=1-0.85z^{-1}$
to the input, and the decoder applies the inverse de-emphasis filter.
This attenuates the low frequencies (LF), reducing the amount of leakage
they cause at higher frequencies (HF). Second, the encoder applies
a perceptual prefilter, with a corresponding postfilter in the decoder,
as Section~\ref{sub:Prefilter-and-Postfilter} describes. 

\begin{figure*}
\includegraphics[width=1\textwidth]{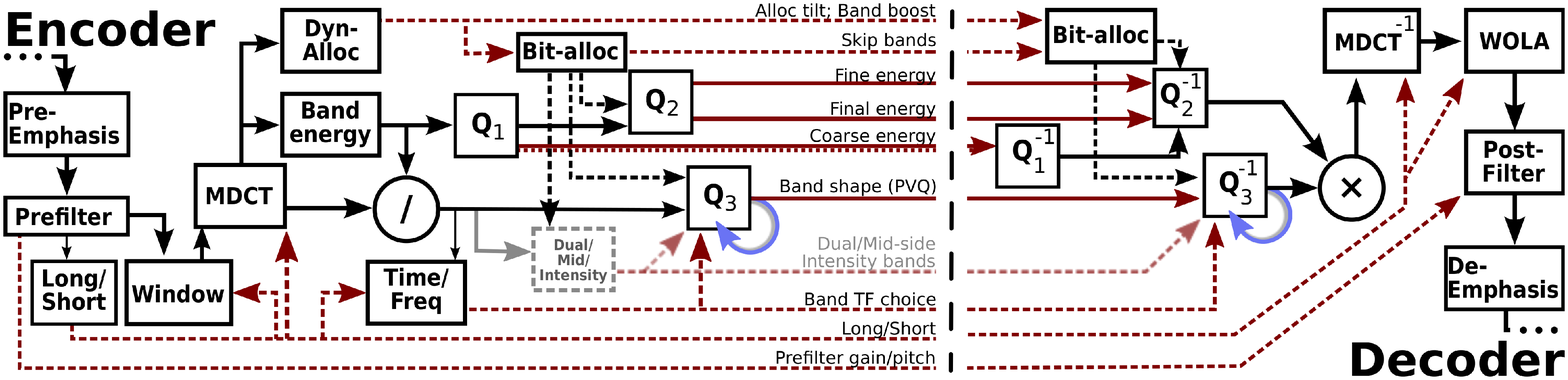}

\caption{Overview of the CELT algorithm.\label{fig:Overview-of-the-CELT-algo}}
\end{figure*}

Fig.~\ref{fig:Overview-of-the-CELT-algo} shows a complete block
diagram of CELT. Sections~\ref{sec:Quantization-and-Encoding} and
\ref{sec:Psychoacoustic-Improvements} describe the various components.

\subsection{Handling of Transients}

\label{sub:Handling-of-Transients}Like other transform codecs, Opus
controls pre-echo primarily by varying the MDCT size. When the encoder
detects a transient, it computes multiple short MDCTs over the frame
and interleaves the output coefficients. For 20\nobreakdash-ms frames,
there are 8~MDCTs with full-overlap, 5~ms windows. We constrain
the band sizes to be a multiple of the number of short MDCTs, so that
the interleaved coefficients form bands of the same size, covering
the same part of the spectrum in each block, as the corresponding
band of a long MDCT.

\section{Quantization and Encoding}

\label{sec:Quantization-and-Encoding}Opus supports any bitrate that
corresponds to an integer number of bytes per frame. Rather than signal
a rate explicitly in the bitstream, Opus relies on the lower-level
transport protocol, such as RTP, to transmit the payload length. The
decoder, not the encoder, makes many bit allocation decisions automatically
based on the number of bits remaining. This means that the encoder
must determine the final rate early in the encoding process, so it
can make matching decisions, unlike codecs such as AAC, MP3, and Vorbis.
This has two advantages. First, the encoder need not transmit these
decisions, avoiding the associated overhead. Second, they allow the
encoder to achieve a target bitrate exactly, without repeated encoding
or bit reservoirs. Even though entropy coding produces variable-sized
output, these dynamic adjustments to the bit allocation ensure that
the coded symbols never exceed the number of bytes allocated for the
frame by the encoder earlier in the process. In the vast majority
of cases, the encoder also wastes less than two bits.

\begin{figure}
\includegraphics[width=1\columnwidth]{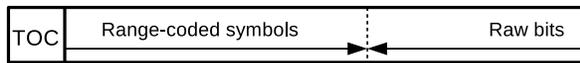}

\caption{Layout and coding order of the bitstream.\label{fig:Layout-and-coding-order}}
\end{figure}

Opus encodes most symbols using a range coder~\cite{Mart79}. Some
symbols, however, have a power-of-two range and approximately uniform
probability. Opus packs these as \emph{raw bits}, starting at the
end of the packet, back towards the end of the range coder output,
as Fig.~\ref{fig:Layout-and-coding-order} illustrates. This allows
the decoder to rapidly switch between decoding symbols with the range
coder and reading raw bits, without interleaving the data in the packet.
It also improves robustness to bit errors, as corruption in the raw
bits does not desynchronize the range coder. A special termination
rule for the range coder, described in Section~5.1.5 of RFC~6716~\cite{rfc6716},
ensures the stream remains decodable regardless of the values of the
raw bits, while using at most 1~bit of padding to separate the two.

\subsection{Coarse Energy Quantization ($Q_{1}$)}

The most important information encoded in the bitstream is the energy
of the MDCT coefficients in each band. Band energy is quantized using
a two-pass coarse-fine quantizer. The coarse quantizer uses a fixed
6~dB resolution for all bands, with inter-band prediction and, optionally,
inter-frame prediction. The 2D \emph{z}\nobreakdash-transform of
the predictor is
\begin{equation}
A\left(z_{\ell},z_{b}\right)=\left(1-\alpha z_{\ell}^{-1}\right)\cdot\frac{1-z_{b}^{-1}}{1-\beta z_{b}^{-1}}\ ,
\end{equation}
where $\ell$ is the frame index and $b$ is the band. Inter-frame
prediction can be turned on or off for any frame. When enabled, both
$\alpha$ and $\beta$ are non-zero and depend on the frame size.
When disabled, $\alpha=0$ and $\beta=0.15$. Inter-frame prediction
is more efficient, but less robust to packet loss. The encoder can
use packet loss statistics to force inter-frame prediction off adaptively.
The prediction residual is entropy-coded assuming a Laplace probability
distribution with per-band variances trained offline.

\subsection{Bit Allocation}

\label{sub:Bit-Allocation}
\begin{figure}
\begin{center}\includegraphics[width=0.81\columnwidth]{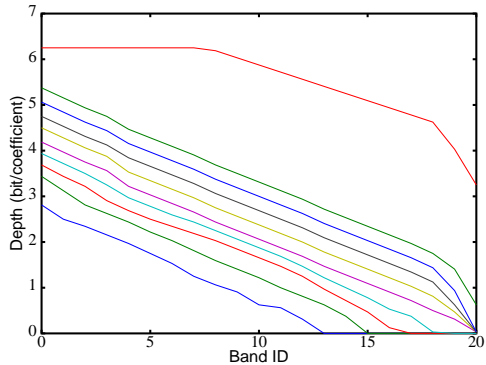}\end{center}

\caption{Static bit allocation curves in bits/sample for each band and for
multiple bitrates.\label{fig:Static-bit-allocation}}
\end{figure}

Rather than transmitting scale factors
like MP3 and AAC or a floor curve like Vorbis, CELT mostly allocates
bits implicitly. After coarse energy quantization, the encoder decides
on the total number of bytes to use for the frame. Then both the encoder
and decoder run the same \emph{bit-exact} bit allocation function
to partition the bits among the bands. CELT interpolates between several
static allocation prototypes (see Fig.~\ref{fig:Static-bit-allocation})
to achieve the target rate.

Some bands may not receive any bits. The decoder reconstructs them
using only the energy, generating fine details by spectral folding,
as Section~\ref{sub:Spectral-Folding} details. When a band receives
very few bits, the sparse spectrum that could be encoded with them
would sound worse than spectral folding. Such bands are automatically
skipped, redistributing the bits they would have used to code their
spectrum to the remaining bands. The encoder can also skip more bands
via explicit signaling. This allows it to give the skip decisions
some hysteresis between frames. 

After the initial allocation, bands are encoded one at a time. In
practice, a band may use slightly more or slightly fewer bits than
allocated. The difference propagates to subsequent bands to ensure
that the final rate still matches the overall target. Automatically
adjusting the allocation based on the actual bits used makes achieving
CBR easy. 

The implicit allocation produces a nearly constant signal-to-noise
ratio in each band, with the LF coded at a higher resolution than
the HF. It approximates the real masking curve well without any signaling,
and achieves good quality by itself, as demonstrated by earlier versions
of the algorithm~\cite{Valin2009,Valin2010}. However, it does not
cover two theoretical phenomena:
\begin{enumerate}
\item Tonality: tones provide weaker masking than noise, requiring a finer
resolution. Since tones usually have harmonics in many bands, the
encoder increases the total rate for these frames.
\item Inter-band masking: a band may be masked by neighboring bands, though
this is weaker than intra-band masking. 
\end{enumerate}
CELT provides two signaling mechanisms that adjust the implicit allocation:
one that changes the tilt of the allocation, and one that boosts specific
bands.

\subsubsection{Allocation Tilt}

The allocation tilt parameter changes the slope of the bit allocation
as a function of the band index by up to $\pm5/64$~bit/sample/band,
in increments of $1/64$~bit/sample/band. Although in theory the
slope of the masking threshold should follow the slope of the signal's
spectral envelope, we have observed that LF-dominated signals require
more bits in the LF, with a similar observation for HF-dominated signals.

\subsubsection{Band Boost}

When a specific band requires more bits, the bitstream includes a
mechanism for increasing its allocation (reducing the allocation of
all other bands). Versions 1.0.x and earlier of the Opus reference
implementation rarely use this band boost. However, newer versions
use it to improve quality in the following circumstances:
\begin{itemize}
\item \itemspace In transients frames, bands dominated by the leakage of
the shorter MDCTs receive more bits.
\item \itemspace Bands that have significantly larger energy than surrounding
bands receive more bits.
\end{itemize}
\itemspace CELT does not provide a mechanism to reduce the allocation
of a single band because it would not be worth the signaling cost.

\subsection{Fine Energy Quantization ($Q_{2}$)}

Once the per-band bit allocation is determined, the encoder refines
the coarsely-quantized energy of each band. Let $a$ be the total
allocation for a band containing $N_{DoF}$ degrees of freedom\footnote{Usually equal to the number of coefficients. When stereo coupling
is used on a band with more than 2 coefficients, the combined band
has an additional degree of freedom.}. We approximate (30) from \cite{Kruger2008} to obtain the fine energy
allocation:
\begin{equation}
a_{f}=\frac{a}{N_{DoF}}+\frac{1}{2}\log_{2}N_{DoF}-K_{fine}\ ,
\end{equation}
where $K_{fine}$ is a tuned \emph{fine allocation offset}. We round
the result to an integer and code the refinement data as raw bits.
Bands where $N_{DoF}=2$ get slightly more bits, and we slightly bias
the allocation upwards when adding the first and second fine bit. 

If any bits are left unused at the very end of the frame, each band
may add one additional bit per channel to refine the band energy,
starting with bands for which $a_{f}$ was rounded down.

\subsection{Pyramid Vector Quantization ($Q_{3}$)}

Let $\mathbf{X}_{b}$ be the MDCT coefficients for band $b$. We normalize
the band with the unquantized energy,
\begin{equation}
\mathbf{x}_{b}=\frac{\mathbf{X}_{b}}{\left\Vert \mathbf{X}_{b}\right\Vert }\ ,
\end{equation}
producing a unit vector on an $N$\nobreakdash-sphere, coded with
a pyramid vector quantizer (PVQ)~\cite{Fisher1986} codebook:
\[
S\left(N,K\right)=\left\{ \frac{\mathbf{y}}{\left\Vert \mathbf{y}\right\Vert },\ \mathbf{y}\in\left\{ \mathbb{Z^{N}}:\sum_{i=0}^{N-1}\left|y_{i}\right|=K\right\} \right\} \ ,
\]
where $K$ is the $L_{1}$\nobreakdash-norm of $\mathbf{y}$, i.e.
the number of \emph{pulses}. The codebook size obeys the recurrence
\begin{multline}
V\left(N,K\right)=V\left(N,K-1\right)\\
+V\left(N-1,K\right)+V\left(N-1,K-1\right)\ ,
\end{multline}
with $V\left(N,0\right)=1$ and $V\left(0,K\right)=0,\ K>0$. Because
$V\left(N,K\right)$ is rarely a power of two, we use the range coder
with a uniform probability to encode the codeword index, derived from
$\mathbf{y}$ using the method of~\cite{Fisher1986}. When $V\left(N,K\right)$
is larger than 255, the index is renormalized to fall in the range
$\left[128,255\right]$ and the least significant bits are coded using
raw bits. The uniform probability allows both the encoder and the
decoder to choose $K$ such that $\log_{2}V\left(N,K\right)$ achieves
allocation determined in Section~\ref{sub:Bit-Allocation}.

\subsubsection{Spectral Folding}

\label{sub:Spectral-Folding}When a band receives no bits, the decoder
replaces the spectrum of that band with a normalized copy of MDCT
coefficients from lower frequencies. This preserves some temporal
and tonal characteristics from the original band, and CELT's energy
normalization preserves the spectral envelope. Spectral folding is
far less advanced than spectral band replication (SBR) from HE-AAC
and mp3PRO, but is computationally inexpensive, requires no extra
delay, and the decision to apply it can change frame-by-frame.

\subsection{Stereo}

Opus supports three different stereo coupling modes:
\begin{enumerate}
\item \itemspace Mid-side (MS) stereo
\item \itemspace Dual stereo
\item \itemspace Intensity stereo
\end{enumerate}
\itemspace A coded band index denotes where intensity stereo begins:
all bands above it use intensity stereo, while all bands below it
use either MS or dual stereo. A single flag at the frame level chooses
between them.

\subsubsection{Mid-Side Stereo}

We apply MS stereo coupling separately on each band, after normalization.
Because we code the energy of each channel separately, MS stereo coupling
never introduces cross-talk between channels and is safe even when
dual stereo is more efficient. Let $\mathbf{x}_{l}$ and $\mathbf{x}_{r}$
be the normalized band for the left and right channels, respectively.
The orthogonal mid and side signals are computed as
\begin{align}
\mathbf{M} & =\frac{\mathbf{x}_{l}+\mathbf{x}_{r}}{2}\ ,\\
\mathbf{S} & =\frac{\mathbf{x}_{l}-\mathbf{x}_{r}}{2}\ .
\end{align}

Opus encodes the mid and side as normalized signals $\mathbf{m}=\mathbf{M}/\left\Vert \mathbf{M}\right\Vert $
and $\mathbf{s}=\mathbf{S}/\left\Vert \mathbf{S}\right\Vert $. To
recover $\mathbf{M}$ and $\mathbf{S}$ from $\mathbf{m}$ and $\mathbf{s}$,
we need to know the ratio of $\left\Vert \mathbf{S}\right\Vert $
to $\left\Vert \mathbf{M}\right\Vert $, which we encode as the angle
\begin{equation}
\theta_{s}=\arctan\frac{\left\Vert \mathbf{S}\right\Vert }{\left\Vert \mathbf{M}\right\Vert }\ .\label{eq:stereo-angle}
\end{equation}
Explicitly coding $\theta_{s}$ preserves the stereo width and reduces
the risk of \emph{stereo unmasking}~\cite{Johnston1992,Moore}, since
it preserves the energy of the difference signal, in addition to the
energy in each channel. We quantize $\theta_{s}$ uniformly, deriving
the resolution the same way as the fine energy allocation. Uniform
quantization of $\theta_{s}$ achieves optimal mean-squared error
(MSE).

Let $\hat{\mathbf{m}}$ and $\hat{\mathbf{s}}$ be the quantized versions
of $\mathbf{m}$ and $\mathbf{s}$. We can compute the reconstructed
signals as
\begin{align}
\hat{\mathbf{x}}_{l} & =\hat{\mathbf{m}}\cos\hat{\theta}_{s}+\hat{\mathbf{s}}\sin\hat{\theta}_{s}\ ,\\
\hat{\mathbf{x}}_{r} & =\hat{\mathbf{m}}\cos\hat{\theta}_{s}-\hat{\mathbf{s}}\sin\hat{\theta}_{s}\ .
\end{align}
As a result of the quantization, $\hat{\mathbf{m}}$ and $\hat{\mathbf{s}}$
may not be orthogonal, so $\hat{\mathbf{x}}_{l}$ and $\hat{\mathbf{x}}_{r}$
may not have exactly unit norm and must be renormalized. 

The MSE-optimal bit allocation for $\mathbf{m}$ and $\mathbf{s}$
depends on $\hat{\theta}_{s}$. Let $N$ be the size of the band and
$a$ be the total number of bits available for $\mathbf{m}$ and $\mathbf{s}$.
Then the optimal allocation for $\mathbf{m}$ is
\begin{equation}
a_{mid}=\frac{a-\left(N-1\right)\log_{2}\tan\theta_{s}}{2}\ .
\end{equation}
The larger of $\mathbf{m}$ and $\mathbf{s}$ is coded first, and
any unused bits are given to the other channel. As a special case,
when $N=2$ we use the orthogonality of $\mathbf{m}$ and $\mathbf{s}$
to code one of the channels using a single sign bit.

\subsubsection{Dual Stereo}

Dual stereo codes the normalized left and right channels independently.
We use this only when the correlation between the channels is not
strong enough to make up for the cost of coding the $\theta_{s}$
angles.

\subsubsection{Intensity Stereo}

Intensity stereo also works in the normalized domain, using a single
mid channel with no side. Instead of $\theta_{s}$, we code a single
\emph{inversion} flag for each band. When set, we invert the right
channel, producing two channels 180~degrees out of phase.

\subsection{Band splitting}

At high bitrates, we allocate some bands hundreds of bits. To avoid
arithmetic on large integers in the PVQ index calculations, we split
bands with more than 32~bits, using the same process as MS stereo.
$\mathbf{M}$ and $\mathbf{S}$ are set to the first and second half
of the band, with $\theta_{s}$ indicating the distribution of energy
between the two halves. If a band contains data from multiple short
MDCTs, we bias the bit allocation to account for pre-echo or forward
masking using $\hat{\theta}_{s}$. If one sub-vector still requires
more than 32~bits, we split it recursively. This recursion stops
after 4~levels (1/16th the size of the original band), which puts
a hard limit on the number of bits a band can use. This limit lies
far beyond the rate needed to achieve transparency in even the most
difficult samples.

\section{Psychoacoustic Improvements}

\label{sec:Psychoacoustic-Improvements}We can achieve good audio
quality using just the algorithms described above. However, four different
psychoacoustically-motivated improvements make coding artifacts even
less audible.

\subsection{Prefilter and Postfilter}

\label{sub:Prefilter-and-Postfilter}
\begin{figure}
\centering{\includegraphics[width=0.85\columnwidth]{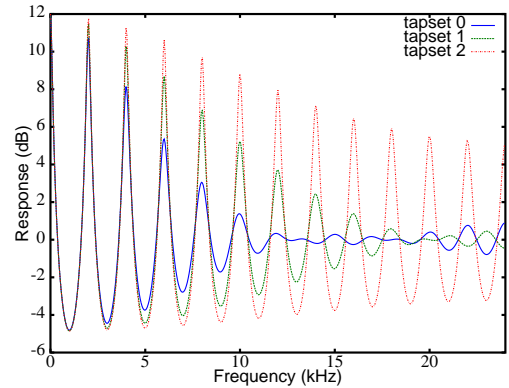}}

\caption{Frequency response of the different postfilter tapsets for $T=24$,
$g=0.75$.\label{fig:Postfilter-response} }
\end{figure}

The low-overlap window increases
leakage in the MDCT, resulting in higher quantization noise on highly
tonal signals. Widely-spaced harmonics in periodic signals provide
especially little masking. Opus mitigates this problem using a pitch-enhancing
post-filter. Unlike speech codec postfilters, we run a matching prefilter
on the encoder side. The pair provides perfect reconstruction (in
the absence of quantization), allowing us to enable the postfilter
even at high bitrates. Although the filters look like a pitch predictor,
unlike standard pitch prediction we apply the prefilter to the unquantized
signal, allowing pitch periods shorter than the frame size. The gain
and period are transmitted explicitly. When these change between two
frames, the filter response is interpolated using a 2.5~ms cross-fade
window equal to the square of the $w(n)$ power-complementary window.
We use a 5-tap prefilter with an impulse response of
\begin{multline}
A\left(z\right)=1-g\cdot\left[a_{p,2}\left(z^{-T-2}+z^{-T+2}\right)\right.\\
\left.{}+a_{p,1}\left(z^{-T-1}+z^{-T+1}\right)+a_{p,0}z^{-T}\right]\ ,
\end{multline}
where $T$ is the pitch period, $g$ is the gain, and $a_{p,i}$ are
the coefficients of tapset $p$. We choose one of three different
tapsets to control the range of frequencies to which we apply the
enhancement. They are
\begin{equation}
\begin{array}{lcl}
a_{0,\cdot} & = & \left[0.80\;0.10\;0\right]\ ,\\
a_{1,\cdot} & = & \left[0.46\;0.27\;0\right]\ ,\\
a_{2,\cdot} & = & \left[0.30\;0.22\;0.13\right]\ .
\end{array}
\end{equation}
The pitch period lies in the range $\left[15,1022\right]$, and the
gain varies between 0.09 and 0.75. Fig.~\ref{fig:Postfilter-response}
shows the frequency response of each tapset for a period of $T=24$
(2~kHz) and a gain $g=0.75$.

Subjective testing conducted by Broadcom on an earlier version of
the algorithm demonstrated the postfilter's effectiveness~\cite{Broadcom-testing}.

\subsection{Variable Time-Frequency Resolution}

\label{sub:Variable-Time-Frequency-Resolution}Some frames contain
both tones and transients, requiring both good time resolution and
good frequency resolution. Opus achieves this by selectively modifying
the time-frequency (TF) resolution in each band. For example, Opus
can have good frequency resolution for LF tonal content while retaining
good time resolution for a transient's HF. We change the TF resolution
with a Hadamard transform, a cheap approximation of the DCT. When
using multiple short MDCTs (good time resolution), we increase the
frequency resolution of a band by applying the Hadamard transform
to the same coefficient across multiple MDCTs. This can increase the
frequency resolution by a factor of 2 to 8, decreasing the time resolution
by the same amount.

\begin{figure}
\includegraphics[width=0.5\columnwidth]{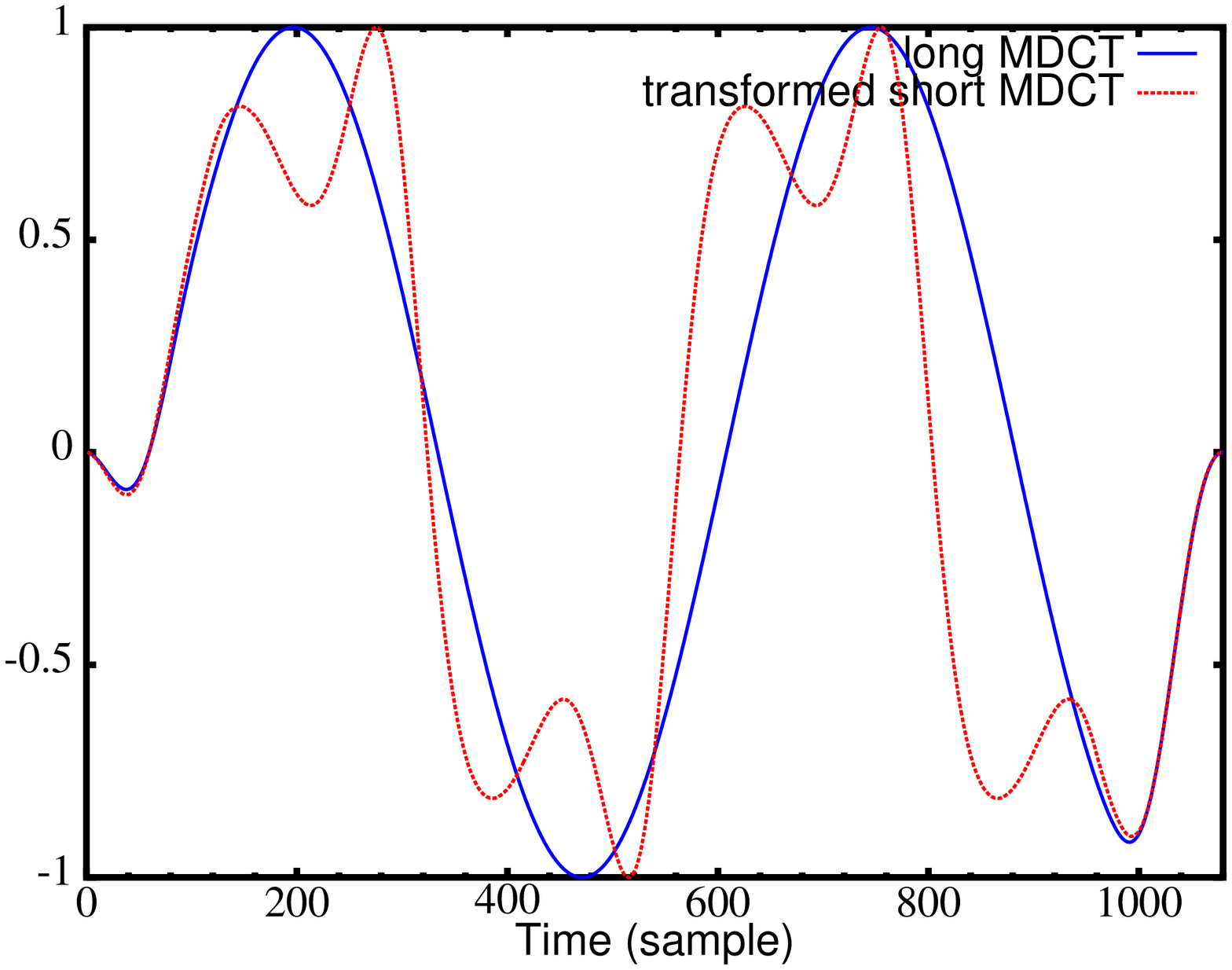}\includegraphics[width=0.5\columnwidth]{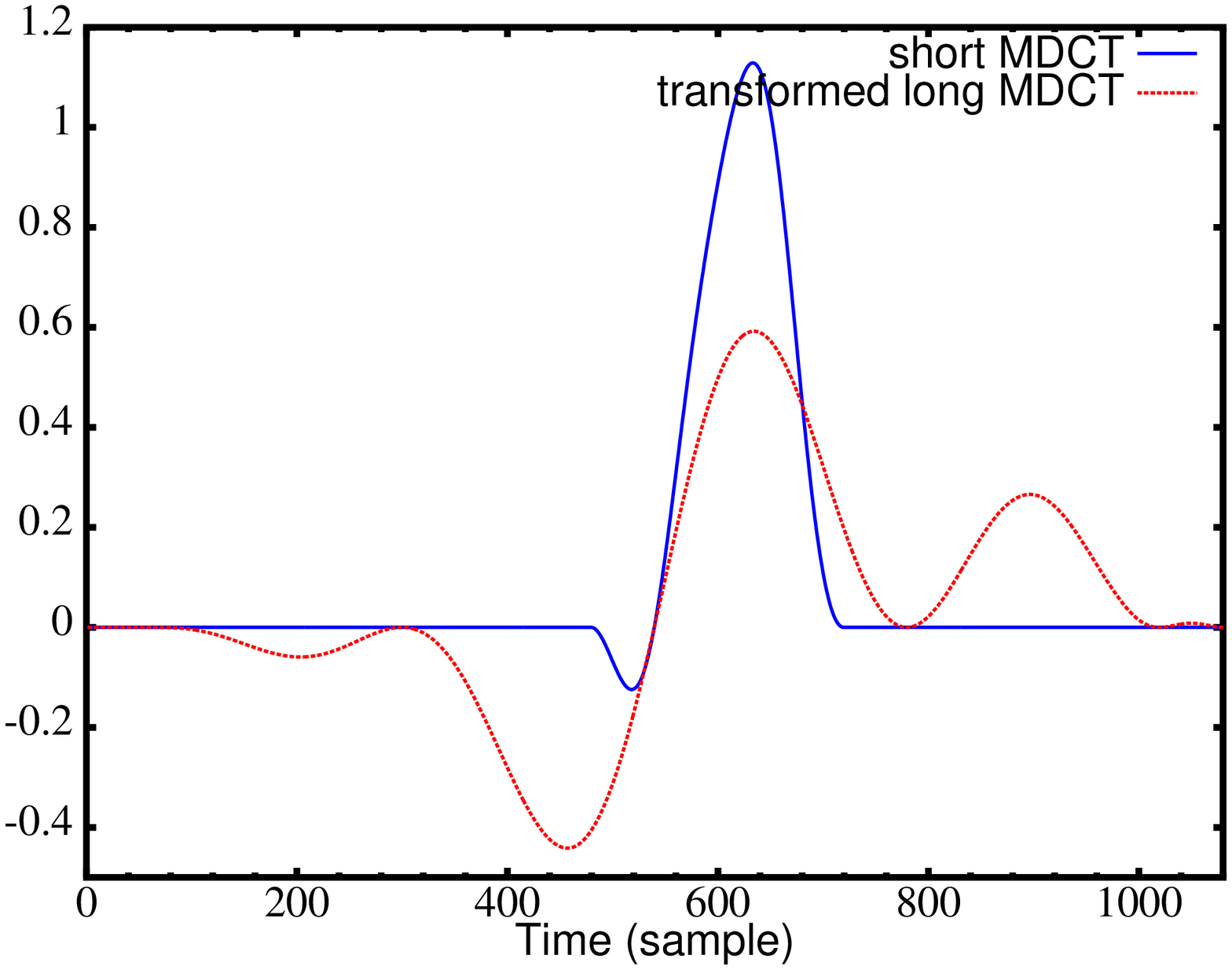}

\caption{Basis functions with modified time-frequency resolution for a 20~ms
frame. Left: fourth basis function of a long MDCT vs. the equivalent
basis function from TF modification of 8 short MDCTs. Right: First
(``DC'') basis function of a short MDCT vs. the equivalent basis
function from TF modification of the first 8 coefficients of a long
MDCT.\label{fig:TF-modified-basis-functions}}
\end{figure}

The Hadamard transform of consecutive coefficients increases the time
resolution of a long MDCT. This yields more time-localized basis functions,
although they have more ringing than the equivalent short MDCT basis
functions. Fig.~\ref{fig:TF-modified-basis-functions} illustrates
basis functions produced by adaptively modifying the time-frequency
resolution for a 20~ms frame. 

\subsection{Spreading Rotations}

\begin{figure}
\includegraphics[width=1\columnwidth]{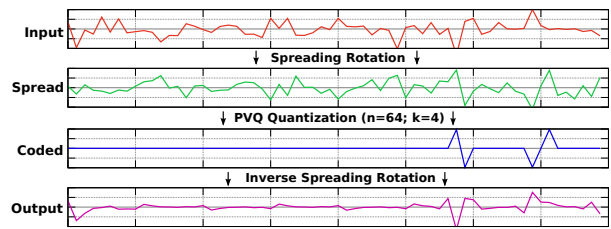}

\caption{Spreading example.\label{fig:Spreading-example}}
\end{figure}

A common type of artifact in transform codecs is tonal noise, also
known as \emph{birdies}. When quantizing a large number of HF MDCT
coefficients to zero, the few remaining non-zero coefficients sound
tonal even when the original signal did not. This is most noticeable
in low-bitrate MP3s. Opus greatly reduces tonal noise by applying
\emph{spreading rotations}. The encoder applies these rotations to
the normalized signal prior to quantization, and the decoder applies
the inverse rotations, as Fig.~\ref{fig:Spreading-example} shows. 

We construct the spreading rotations from a series of 2D Givens rotations.
Let $\mathbf{G}\left(m,n,\theta_{r}\right)$ denote a Givens rotation
matrix by angle $\theta_{r}$ between coefficients $m$ and $n$ in
some band with $N$ coefficients, with angles near $\pi/4$ implying
more spreading. Then the spreading rotations are
\begin{multline}
\mathbf{R}\left(\theta_{r}\right)=\prod_{k=0}^{N-3}\mathbf{G}\left(k,k+1,\theta_{r}\right)\\
\cdot\prod_{k=2}^{N}\mathbf{G}\left(N-k,N-k+1,\theta_{r}\right)\ .\label{eq:Givens-combination}
\end{multline}
In other words, we rotate adjacent coefficient pairs one at a time
from the beginning of the vector to the end, and then back. We determine
$\theta_{r}$ from the band size, $N$, and the number of pulses used,
$K$: 
\begin{equation}
\theta_{r}=\frac{\pi}{4}\left(\frac{N}{N+\delta K}\right)^{2}\ ,\label{eq:Theta-N-K}
\end{equation}
where $\delta$ is the \emph{spreading constant}. Once per frame,
the encoder selects $\delta$ from one of three values: 5, 10, or
15, or disables spreading completely.

In transient frames, we apply the spreading rotations to each short
MDCT separately to avoid pre-echo. When vectors of more than 8 coefficients
need to be rotated, we apply an additional set of rotations to pairs
of coefficients $\bigl\lfloor\sqrt{N}\bigr\rfloor$ positions apart,
using the angle $\theta_{r}^{'}=\frac{\pi}{2}-\theta_{r}$. This spreads
the energy within large bands more widely. 

\subsection{Collapse Prevention}

\begin{figure}
\centering{\includegraphics[width=0.96\columnwidth,height=0.54\columnwidth]{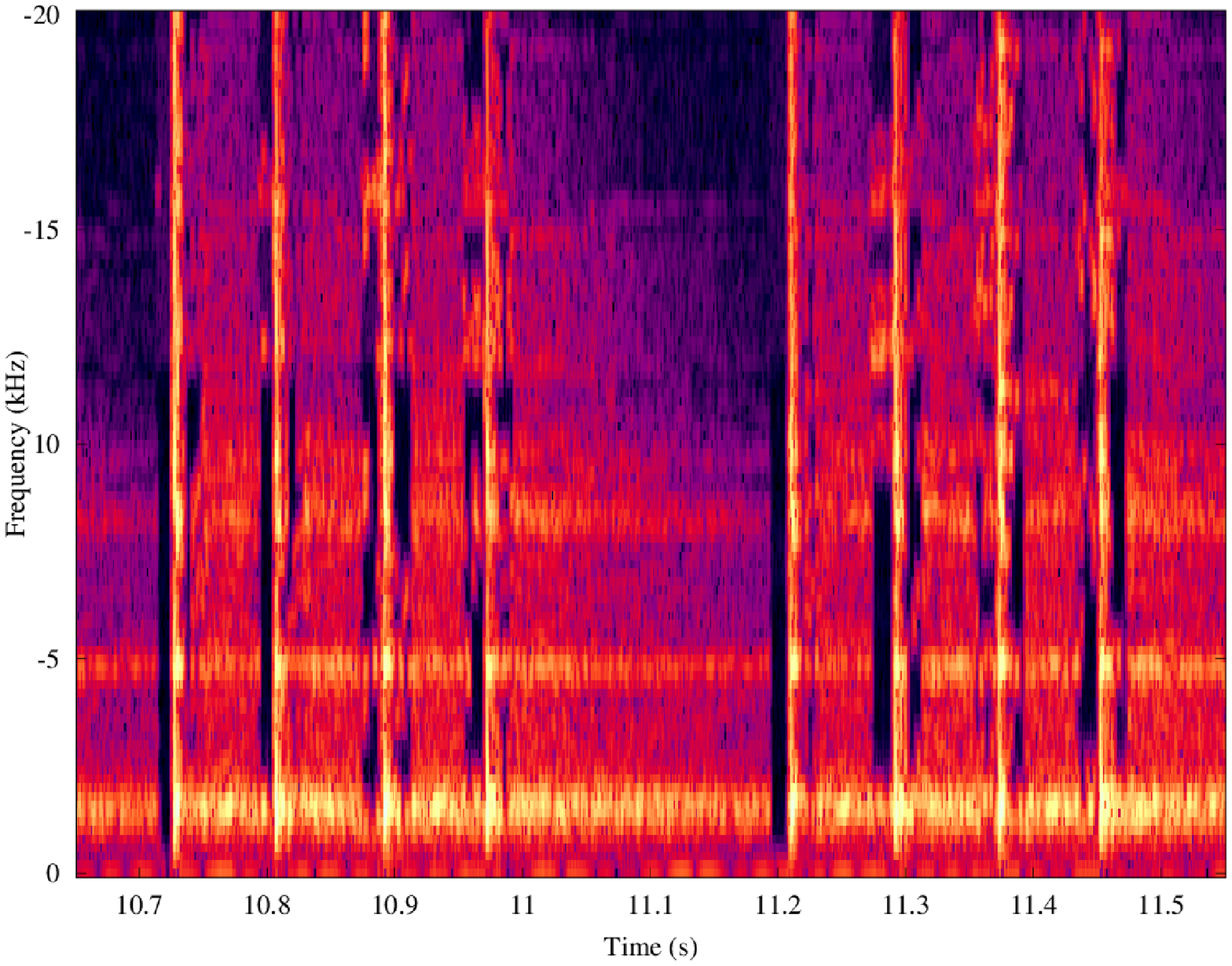}}

\centering{\includegraphics[width=0.96\columnwidth,height=0.54\columnwidth]{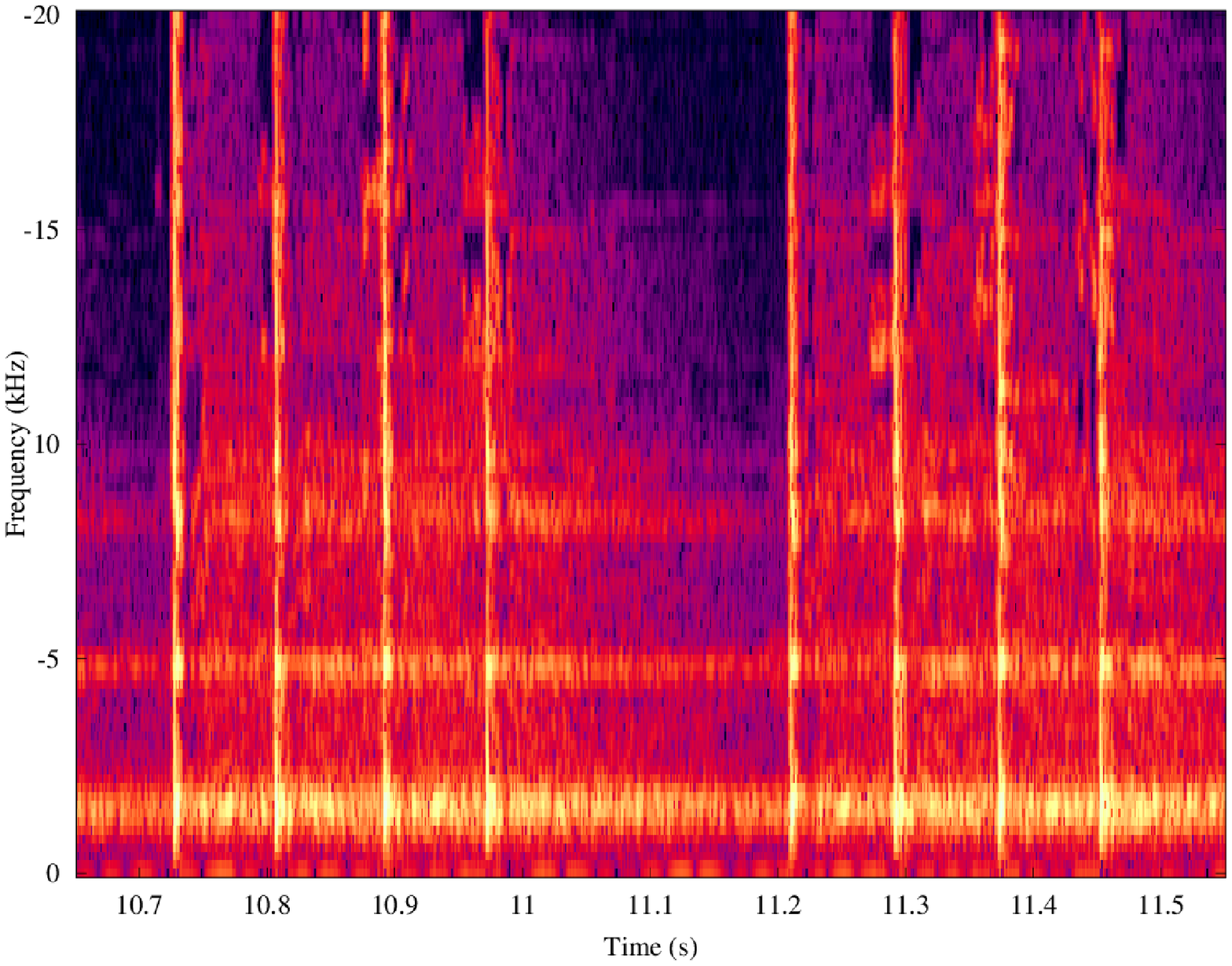}}

\caption{Extreme collapse prevention example for castanets at 32~kb/s mono.
Top: without collapse prevention. Bottom: with collapse prevention.\label{fig:Collapse-prevention-example}}
\end{figure}

In transients at low bitrates, Opus may quantize all of the coefficients
in a band corresponding to a particular short MDCT to zero. Even though
we preserve the energy of the entire band, this quantization causes
audible drop outs, as Fig.~\ref{fig:Collapse-prevention-example}
shows on the left. The decoder detects \emph{holes} that occur when
a short MDCT receives no pulses in a given band, or when folding copies
such a hole into a higher band, and fills them with pseudo-random
noise at a level equal to the minimum band energy over the previous
two frames. The encoder transmits one flag per frame that can disable
collapse prevention. We do this after two consecutive transients to
avoid putting too much energy in the holes. Fig.~\ref{fig:Collapse-prevention-example}
shows the result of collapse prevention on the right. The short drop
outs around each transient are no longer audible.

\section{Evaluation and Results}

\label{sec:Evaluation-and-Results}This section presents a quality
evaluation of Opus's CELT mode on music signals. More complete evaluation
data on Opus is available at~\cite{Opus-results}.

\subsection{Subjective Quality}

\begin{figure}
\centering{\includegraphics[width=0.98\columnwidth]{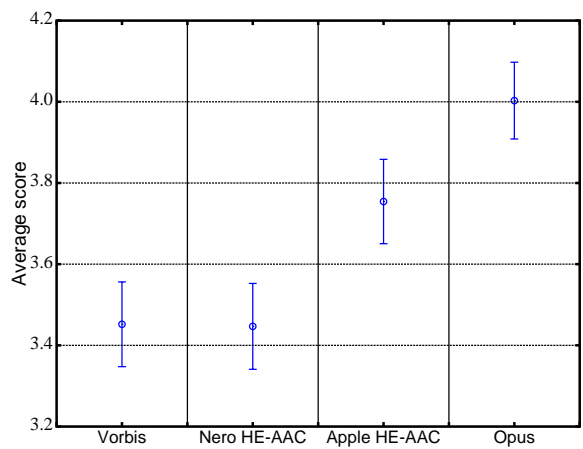}}\caption{Results of the 64~kb/s evaluation. The low anchor (omitted) was rated
at 1.54 on average.\label{fig:Results-of-the-64kbps-evaluation}}
\end{figure}

Volunteers of the HydrogenAudio forum\footnote{http://hydrogenaudio.org/}
evaluated the quality of 64~kb/s VBR Opus on fullband stereo music
with headphones. 13~listeners evaluated 30~samples using the ITU-R
BS.1116-1 methodology~\cite{BS1116-1} with
\begin{itemize}
\item \itemspace The Opus~\cite{Opus-website} reference implementation
(v0.9.2),
\item \itemspace Apple's HE-AAC\footnote{With constrained VBR, as it cannot run unconstrained }
(QuickTime v7.6.9),
\item \itemspace Nero's HE-AAC\footnote{http://www.nero.com/enu/company/about-nero/nero-aac-codec.php}
(v1.5.4.0), and
\item \itemspace Ogg Vorbis (AoTuV\footnote{http://www.geocities.jp/aoyoume/aotuv/}
v6.02 Beta).
\end{itemize}
\itemspace Apple's AAC-LC at 48~kb/s served as a low anchor. Fig.~\ref{fig:Results-of-the-64kbps-evaluation}
shows the results. A pairwise resampling-based free step-down analysis
using the max(T) algorithm~\cite{Westfall93,Pascutto11} reveals
that Opus is better than the other codecs with greater than 99.9\%
confidence. Apple's HE-AAC was better than both Nero's HE-AAC and
Vorbis with greater than 99.9\% confidence. Nero's HE-AAC and Vorbis
were statistically tied. A simple ANOVA analysis gives the same results.

\subsection{Cascading Performance}

\begin{figure}
\includegraphics[width=0.5\columnwidth]{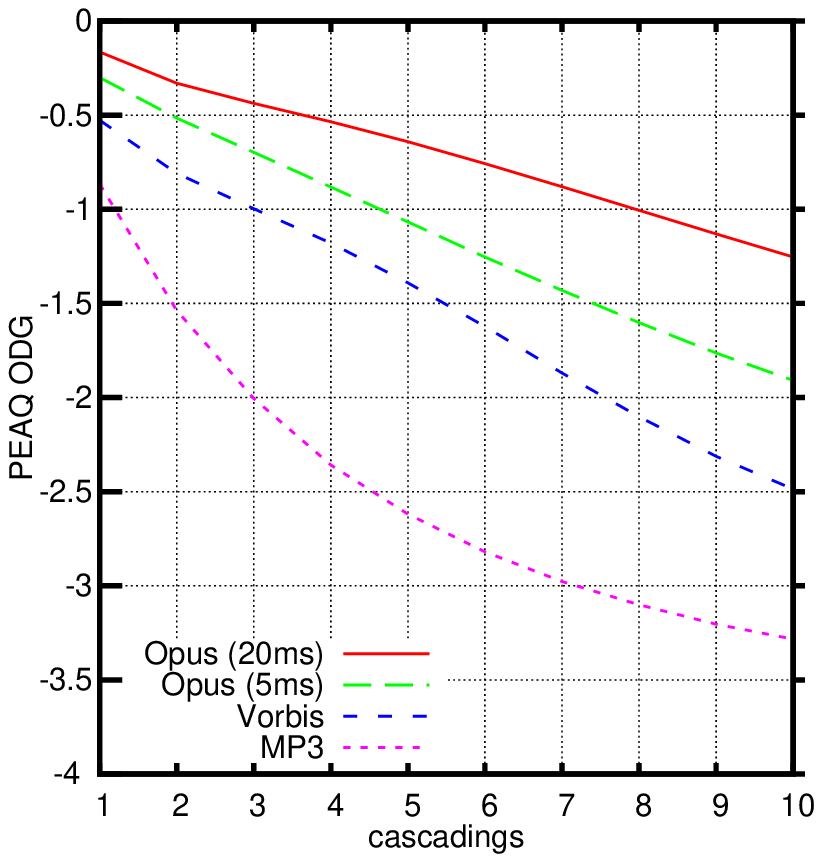}\includegraphics[width=0.5\columnwidth]{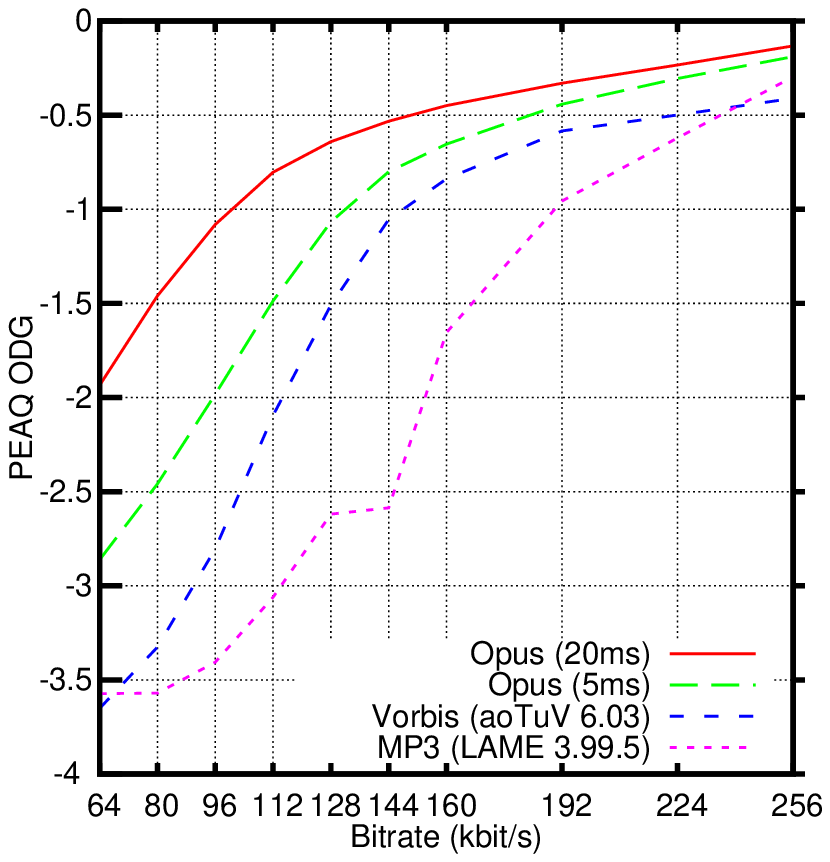}\caption{Cascading quality. Left: Quality degradation vs. number of cascadings
at 128~kb/s. Right: Quality degradation vs. bitrate after 5~cascadings.
\label{fig:Cascading-quality-results} }
\end{figure}

In broadcasting applications, audio streams are compressed and recompressed
multiple times. According to~\cite{Marston2005}, typical broadcast
chains may include up to 5 lossy encoding stages. For this reason,
we compare the cascading quality of Opus to both Vorbis and MP3 using
PQevalAudio~\cite{PQevalAudio}, an implementation of the PEAQ basic
model~\cite{BS1387}. Fig~\ref{fig:Cascading-quality-results} plots
quality as a function of bitrate and the number of cascaded encodings.
Opus performs better than MP3 and Vorbis in the presence of cascading,
with 64~kb/s Opus even out-performing 128~kb/s MP3. Although the
Opus quality with 5~ms frames is lower than for 20~ms frames, it
is still acceptable, and better than MP3. 

\section{Conclusion and Future Work}

By building psychoacoustic knowledge into the Opus format, we minimize
the side information it transmits and the impact of coding artifacts.
This allows Opus to achieve higher music quality than existing non-real-time
codecs, even under cascading. Since Opus was only recently standardized,
we are continuing to improve its encoder, experimenting with such
things as look-ahead and automatic frame size switching for non-real-time
encoding.

\bibliographystyle{unsrt}
\bibliography{celt}

\end{document}